# InteractionNet: Modeling and Explaining of Noncovalent Protein-Ligand Interactions with Noncovalent Graph Neural Network and Layer-Wise Relevance Propagation

Hyeoncheol Cho[1], Eok Kyun Lee[1*], and Insung S. Choi[1*]

[1]Department of Chemistry, KAIST, Daejeon 34141, Korea.
Email: eklee@kaist.ac.kr, ischoi@kaist.ac.kr.

**Abstract:** Expanding the scope of graph-based, deep-learning models to noncovalent protein-ligand interactions has earned increasing attention in structure-based drug design. Modeling the protein-ligand interactions with graph neural networks (GNNs) has experienced difficulties in the conversion of protein-ligand complex structures into the graph representation and left questions regarding whether the trained models properly learn the appropriate noncovalent interactions. Here, we proposed a GNN architecture, denoted as InteractionNet, which learns two separated molecular graphs, being covalent and noncovalent, through distinct convolution layers. We also analyzed the InteractionNet model with an explainability technique, i.e., layer-wise relevance propagation, for examination of the chemical relevance of the model's predictions. Separation of the covalent and noncovalent convolutional steps made it possible to evaluate the contribution of each step independently and analyze the graph-building strategy for noncovalent interactions. We applied InteractionNet to the prediction of protein-ligand binding affinity and showed that our model successfully predicted the noncovalent interactions in both performance and relevance in chemical interpretation.

# Introduction

Deep-learning chemistry is an emerging field in the chemistry discipline, and it has shown remarkable fruition in diverse chemical areas.[1-4] It is the representation learning of molecules, without the human-curated heuristics and descriptors, used widely in cheminformatics for decades, which is one of the recent endeavors of deep-learning chemistry. Learning of the representations from actual molecular structures would enable the full utilization of the discriminative power of deep neural networks (DNNs) on the prediction of target properties without prior quantum-chemical calculations.[5,6] Compared with the physics-based computational methods for calculating molecular properties, the deep-learning approach offers a fast, but still powerful, option for estimating diverse characteristics of molecules through the data-driven discovery of molecular patterns.[7,8]

The recent rise of graph neural networks (GNNs) has upscaled deep-learning capability in chemistry with the easy handling of molecules as molecular graphs, which are defined by two sets of vertices and edges.[9,10] The molecular graphs contain the structural information on molecules in two-dimensional (2D) space, with atoms as vertices and bonds as edges in the graphs. In the GNN, the neurons in a layer are connected to their graph neighborhoods, and layer stacking generates broader local structures in molecules. Many GNN models have been developed for the prediction of molecular energies,[7,8] physical properties,[11,12] protein interactions,[13,14] and biochemical functions.[15,16] The pioneering reports, by Duvenaud et al.[11] and Kearnes et al.,[15] showed the effectiveness of GNNs on the prediction of molecular properties compared to other machine-learning methods based on molecular fingerprints, and the GNN architecture has further been refined to message-passing neural networks (MPNNs).[7] Moreover, the expansion of the GNN architecture into the 3D space for modeling the actual molecular structures has recently been explored, and the efficacy of the GNN approach on the problems requiring 3D molecular structures has been proven.[14,17]

One of the focused fields of deep learning in chemistry is the replacement of the scoring function on the structure-based drug design with data-driven DNN models.[13,14,18-22] The essence of DNN models for deep-learning scoring compared to the force field energy functions and scoring functions is the appropriate database to learn molecular patterns and their relationship to binding affinity, and they are strongly linked to prediction performance. The PDBbind database[23,24] is the most widely used dataset, which is a curation of 3D protein-ligand structures obtained from X-ray crystallography and multidimensional NMR techniques with complementary binding affinities, for training DNN models on prediction of the binding constant from a complex structure. Many DNN models were developed based on the PDBbind database and can be classified into two categories: convolutional neural networks (CNNs) with voxelized images and GNNs working on graph representations of complexes.

Rapid development of high-performance and deeply-stacked CNN models in computer science has dramatically raised the prediction performance of binding constants through enhanced pattern recognition of the 3D molecular images.[21,22,25,26] Protein-ligand complexes were transferred into the angstrom-level voxel grid and used for training the CNN models. Meanwhile, the GNN models,[13,14] which focused on the interpretation of molecular bonding (i.e., covalent bonds) as graph edges, was utilized after the success of CNN models by incorporating noncovalent (NC) interactions as graph edges in molecular graphs, which play significant roles in the programmed formation of 3D molecular structures of biomolecules (e.g., proteins, nucleic acids, and lipid bilayers) and polymers and their dynamics.[27-29] In the GNN models, NC connectivity was utilized in conjunction with covalent connectivity for post-refinement of atomic features after the convolution with covalent-bond connectivity. Gaussian decay functions have been used to mimic decreased influences from distant atoms, or multiple kernels for distance bins have been adopted for simulating NC interactions.[8,13,30] These approaches enrich the graphic representation of molecules by adding topological information and acquiring the shape-awareness, which has only been feasible in the CNNs. In addition to the decay simulation, an approximation of the entire atomic contribution to a smaller subset was widely utilized.[14,19,20,21] Due to the extremely large number of atoms in the protein-ligand complex compared to other molecules in molecular property datasets, training the complex data is challenging for both CNN and GNN models. By limiting the protein atoms into a spatial neighborhood of the ligand molecule, the complex can be greatly reduced into smaller sizes and trained efficiently without losing important interactions. Owing to the aforementioned advanced approaches, the GNN models became a competitive option for developing deep-learning scoring models with a direct interpretation of molecular structures.

In this paper, we propose a GNN architecture, denoted as InteractionNet, that directly learns molecular graphs without any physical parameters, wherein the NC interactions are encoded as graphs along with the bonded adjacency that models covalent interactions. We utilize the PDBbind dataset for evaluation of the concept and examine the model performance on predicting the binding constant from a complex structure. Specifically, we divide the convolutional layers in InteractionNet into two, the covalent and NC convolution layers ($CV_{[C]}$ and $CV_{[NC]}$ layers, respectively), and evaluate the significance of NC convolution. There have been reports on the incorporation of NC connectivity in GNN models, but in strict combination with covalent connectivity. Here, we apply the covalent and NC connectivity separately to investigate the importance of each convolution layer, which has not been explored. In extreme cases, only $CV_{[NC]}$ is used, without any $CV_{[C]}$ layers, and compared to other models. Moreover, we investigate the optimal cropping strategy for downsizing the protein-ligand structure and efficient training. Based on the findings, we further investigate the explanations for the predictions of the trained model, i.e., how the trained model predicts for the first time from the given input data in the protein-ligand complexation problem. By performing decomposition-based, layer-wise relevance propagation (LRP)[31,32] on behalf of explainable AI[33-35] and visualizing the obtained atomic contribution for the prediction of the protein-ligand complex, we explore the relationship between machine-predicted NC interactions and knowledge-based NC interactions from the molecular structures.

## Results and Discussion

**InteractionNet Architecture.** For graphic representation of a protein-ligand complex, InteractonNet employs two adjacency matrices for the complex, denoted the covalent and NC adjacency matrices ($A_{[C]}$ and $A_{[NC]}$), similar to the PotentialNet reported by Feinberg et al.[13] $A_{[C]}$ and $A_{[NC]}$ are defined by the combination of molecular graphs for a protein and a ligand but with different connectivity strategies. The covalent adjacency matrix, $A_{[C]}$, consists of the bond connectivity in the protein and the ligand, and is constructed by a disjoint union of the protein and the ligand graphs, maintaining the bond connectivity only within each molecule. The NC adjacency matrix, $A_{[NC]}$, defined by a graph having full connectivity between the vertices of the protein and the ligand graphs, contains all the possible edges between the protein and the ligand but not within the same molecule (Figure 1a). Based on the notation used by Feinberg et al., each adjacency matrix can be decomposed into four blocks, $A_{L:L}$, $A_{L:P}$, $A_{P:L}$, and $A_{P:P}$, that correspond to smaller adjacency matrices encoding the connectivity between ligand-ligand, ligand-protein, protein-ligand, and protein-protein atoms, respectively (Equation 1).

$$A = \begin{bmatrix} A_{11} & \cdots & A_{1N} \\ \vdots & \ddots & \vdots \\ A_{N1} & \cdots & A_{NN} \end{bmatrix} = \begin{bmatrix} A_{L:L} & A_{L:P} \\ A_{P:L} & A_{P:P} \end{bmatrix} \quad (1)$$

where $A_{ij}$ is whether the node $i$ and $j$ are adjacent, and $N$ is the number of atoms inside the complex.

For $A_{[C]}$, only the $A_{L:L}$ and $A_{P:P}$ blocks are filled with the existence of a covalent bond between atoms, and the remaining $A_{L:P}$ and $A_{P:L}$ blocks are filled with 0. In the case of $A_{[NC]}$, $A_{L:P}$, and $A_{P:L}$, blocks are filled with 1, implying all possible NC interactions between atoms, and the rest are filled with 0. These adjacency strategies assume that there is no covalent bond between the protein and the ligand, and NC interactions within the same molecule are ignored. Obtained adjacency matrices are used as-is through the neural network without training the adjacency matrix itself or requiring modification during the propagation.

InteractionNet is built to utilize the $A_{[C]}$ and $A_{[NC]}$, consecutively, for the end-to-end prediction of dissociation constants from the molecular structures (Figure 1b). It consists of five functional layers: graph-embedding layers, $CV_{[C]}$ layers, $CV_{[NC]}$ layers, a global pooling (GP) layer, and fully-connected (FC) layers. The graph-embedding layers update the atomic features through fully-connected neural networks that mix the features assigned for each atom. The $CV_{[C]}$ and

CV[NC] layers receive the graph embedding and combine the graph adjacency with atomic features for local aggregation of the information. Compared with the PotentialNet,[13] we separate the spatial convolution layers utilizing the two adjacency matrices, A[C] and A[NC], one-by-one at each layer, whereas they are combined and utilized in a single layer in the PotentialNet. By applying the two adjacency matrices for the spatial convolution separately, we simulate the importance of each step on the prediction of the dissociation constant independently. For each convolution step, InteractionNet utilizes the corresponding adjacency matrix and updates the representation additively by residual connections.[25] After the convolution steps, the GP layer aggregates the atomic features distributed across the atoms in a permutation-invariant way and generates a molecular vector. Sum pooling was utilized for the GP mechanism. With the obtained molecular vector, the FC layers transform the representation into the dissociation constant of a protein-ligand complex.

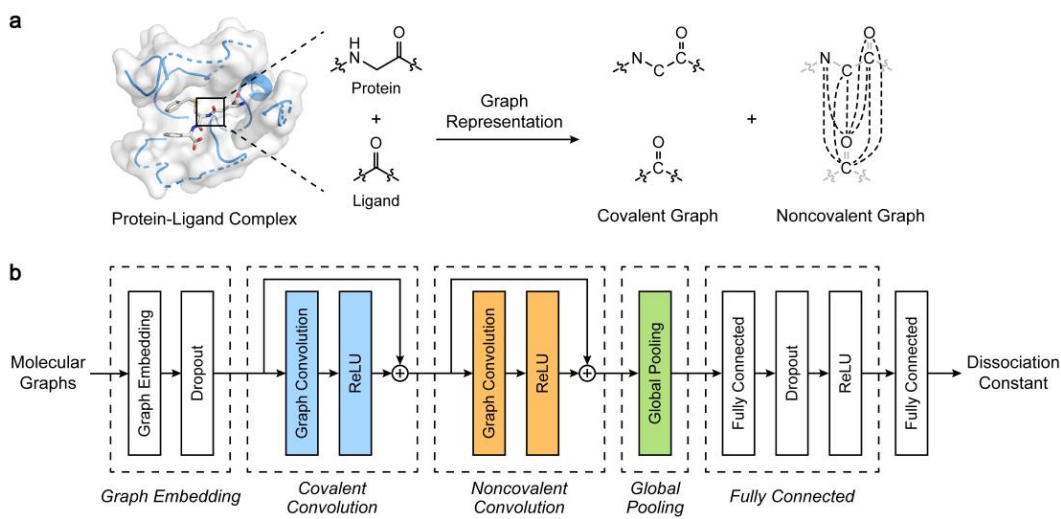

**Figure 1.** (a) Schematic illustrations for modeling NC interactions within a graphic representation, and (b) architecture of InteractionNet for predicting the dissociation constant from the covalent and NC graphs. (a) Structure of the protein-ligand complex converted into two graph representations, encoded by covalent (A[C]) and NC adjacency matrices (A[NC]), defined by covalent bond connectivity and all possible edges between the protein and the ligand, respectively. (b) InteractionNet learns the two aforementioned adjacency matrices, A[C] and A[NC], and predicts the dissociation constant of the complex through a graphic neural network consisting of five functional layers.

**Model Training.** We examined the efficacy of the $CV_{[C]}$ and $CV_{[NC]}$ layers by three variants of InteractionNet with different compositions of CV layers. Four other functional layers of InteractionNet were used with the same number of layers and composition across the variants. By incorporating only one type of the CV layers for InteractionNet, we built InteractionNet$_{[C]}$, utilizing only $CV_{[C]}$ layers, and InteractionNet$_{[NC]}$, utilizing only $CV_{[NC]}$ layers. The variant that incorporated both CV layers sequentially was coined as InteractionNet$_{[C-NC]}$. In the chemists' points of view, InteractionNet$_{[C]}$ focused on the covalent bonds within each ligand and protein molecule for prediction, InteractionNet$_{[NC]}$ did this on NC interactions between the ligand and the protein, and InteractionNet$_{[C-NC]}$ observed covalent bonds first and then used the generated information for the secondary refinement through NC interactions. We chose the dissociation constant of the protein-ligand complex ($K_d$) as our prediction target because the protein-ligand binding is governed primarily by the NC interactions, not covalent bonds, which is important in investigating the efficacy of the proposed architecture. We conducted a 20-fold-cross-validated experiment on the refined set of the PDBbind v2018 dataset,[23,24] consisting of 4,186 complexes and their experimental $K_d$ values.

In the data preprocessing for model training, we cropped the protein structure for faster training and less memory consumption. The binding pockets of the proteins in the refined PDBbind set contained a maximum of 418 atoms, which was almost 16 times larger than the ligands that had only 26 atoms at maximum. We thought that the interactions between a protein and a ligand could be simulated with a smaller subset of atoms in the protein because the number of atoms that participated in the protein-ligand binding is much less than the maximum value. The appropriate cropping strategy, without any loss in performance, is also highly important for efficient training, considering the exponential increase in the memory consumption of the training data. We utilized the spatial atom filtering for simplification of protein structures, which excluded the atoms of a protein distant from a ligand by the range cutoff. In detail, the shortest distance of a protein atom to the ligand atoms was measured, and the protein atom was excluded if the distance exceeded the predefined range cutoff. By spatial cropping with the range cutoff, we obtained a subset of the protein structures, similar to the shape of the van der Waals surface of the ligand but with a much larger radius, and used the subset for the generation of the molecular graphs.

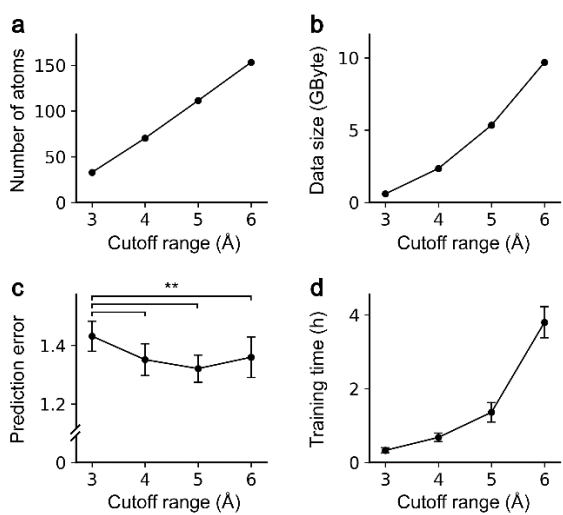

**Figure 2.** Influence of the protein cutoff range from 3 to 6 Å on (a) the average number of atoms included in a complex, (b) the size of the training data, (c) the root-mean-squared-error for the predictions from the trained model, and (d) the average single-fold training time. Error bars indicate standard deviations for each measurement. **$p < 0.005$.

For investigating the influence of the cutoff applied to crop the protein structure into the ligand neighborhoods, we compared the averaged model performance and the training time using InteractionNet[C-NC] by changing the cutoff with 1-Å increment. The number of atoms included in the cropped complex increased linearly with respect to the cutoff, while the data size for the training dataset increased exponentially (Figure 2a-b). The averaged performance was saturated from 4 Å, confirmed by the one-way analysis of variance (ANOVA) with the posthoc Tukey HSD test (Figure 2c). The corresponding training time increased dramatically as the cutoff increased (Figure 2d). Based on the observation, the 5-Å cutoff was considered the most appropriate for our system and used for further investigations.

**Prediction of Dissociation Constants.** The root-mean-square error (RMSE) results from the cross-validation experiments, based on the 5-Å filtering of protein structures, confirmed that the $CV_{[NC]}$ layers played a significant role in the $K_d$ prediction from the molecular graphs (Table 1). InteractionNet$_{[C-NC]}$ and InteractionNet$_{[NC]}$ outperformed InteractionNet$_{[C]}$, regardless of the number of CV layers. For example, the RMSE values for InteractionNet$_{[C-NC]}$ and InteractionNet$_{[C]}$ were 1.321 and 1.379, respectively, showing a 4% improvement by incorporating $CV_{[NC]}$ layers ($p < 0.005$). The performance of InteractionNet$_{[C-NC]}$ was measured to be slightly higher than InteractionNet$_{[NC]}$, but the difference was not significant in statistical analysis ($p = 0.450$). These results indicated that the interactions between a protein and a ligand could be simulated accurately, even with a single $CV_{[NC]}$ layer, without any help from previous covalent-refinement steps.

Table 1. Twenty-fold cross-validation results for InteractionNet on the refined set of the PDBbind v2018. Root-mean-square-errors were measured for each trial and averaged. The best results were highlighted in boldface.

| Model | Train | Validation | Test |
|---|---|---|---|
| InteractionNet$_{[C]}$ | 1.115 ± 0.085 | 1.355 ± 0.060 | 1.379 ± 0.057 |
| InteractionNet$_{[NC]}$ | 1.035 ± 0.077 | 1.328 ± 0.042 | 1.340 ± 0.044 |
| InteractionNet$_{[C-NC]}$ | **0.950 ± 0.032** | **1.313 ± 0.107** | **1.321 ± 0.045** |

For visualization of the individual predictions from the test dataset, we selected the most similar cross-validation trial in performance to the average and obtained a scatterplot and an error histogram between the predicted and experimental values. The scatterplot for the predicted $K_d$ values revealed a high correlation with the experimental $K_d$ in a linear relationship (Figure 3a), and the error distribution showed a Gaussian-like, zero-centered shape (Figure 3b). It is to be noted that 20 cross-validation trials showed similar trends in the scatterplot and the error histogram, but had small differences in pattern (Figure S1).

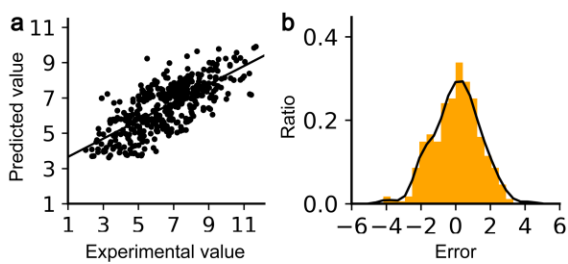

**Figure 3.** (a) Scatterplot and (b) error distribution of predicted and experimental $K_d$ values for 419 complexes included in the test set. (a) The scatterplot for predicted versus experimental $K_d$ is depicted with the trend line (a solid line). (b) The error histogram (orange) and distribution (black) for predictions from the test set. The most similar trial in performance to the average was selected for depicting the graphs in the cross-validation trials.

**Layer-Wise Relevance Propagation (LRP).** The explainability techniques interpret the trained model or their predictions into explanations in human terms, which can be assessed by knowledge-based analysis. By analyzing the system with explainability techniques, the models that fail to learn appropriate knowledge to perform predictions based on valid information and fall into the "Clever Hans" decision made by fragmentary knowledge could be identified.[34] To explore the explainability of the trained InteractionNet model on the $K_d$ prediction, we conducted the post hoc explanation on individual predictions by the LRP.[31,32] The LRP calculates the relevance for every neuron by reversely propagating, through the network, from the predicted output to the input level, and the relevance represents the quantitative contribution of a given neuron to the prediction. We used three LRP rules, LRP-0, LRP-ε, and LRP-γ, sequentially from the output layer to the input layer for production of the relevance for the neurons (Equations 2-4)

$$\text{LRP-0:} \quad R_j = \sum_k \frac{a_j w_{jk}}{\sum_{0,j} a_j w_{jk}} R_k \quad (2)$$

$$\text{LRP-}\epsilon\text{:} \quad R_j = \sum_k \frac{a_j w_{jk}}{\varepsilon + \sum_{0,j} a_j w_{jk}} R_k \quad (3)$$

$$\text{LRP-}\gamma\text{:} \quad R_j = \sum_k \frac{a_j(w_{jk} + \gamma w_{jk}^+)}{\sum_{0,j} a_j(w_{jk} + \gamma w_{jk}^+)} R_k \quad (4)$$

where $j$ and $k$ represent neurons at two consecutive layers, $R$ is the relevance, $a$ denotes lower layer activations, $w^+$ is a positive weight, and $\epsilon$ and $\gamma$ are the parameters used in each LRP rule.

Once we obtained the relevance for the atomic features in the $K_d$ prediction, it was converted to the atomic contributions by summation of relevance for individual features of the same atom. Finally, we compared the relevance with the knowledge-based analysis data from the information on hydrogen bonds and hydrophobic contacts within the complex (Figure 4, see methods for detail). Three protein-ligand complexes from the test set, PDB codes 1KAV, 3F7H, and 4IVB, were sampled and analyzed (Figures 5-7).

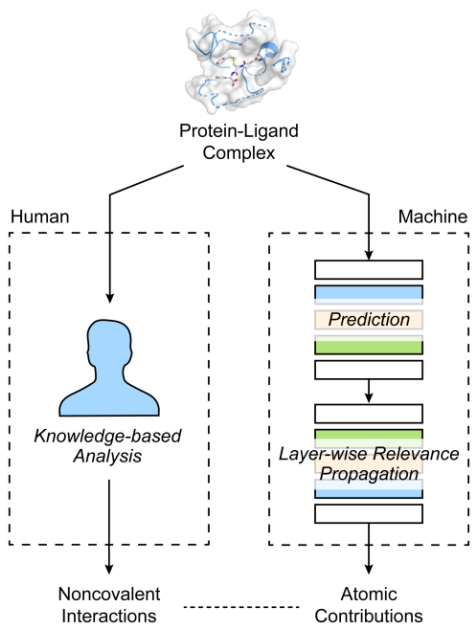

**Figure 4.** Schematic illustration of atomic contributions obtained by applying layer-wise relevance propagation (LRP) on InteractionNet and its comparison with knowledge-based protein-ligand interactions.

*PDB 1KAV: Human tyrosine phosphatase 1B and a phosphotyrosine-mimetic inhibitor (ChEMBL1161222).* As seen in the 3D structure, half of the ligand is surrounded by the protein pocket with substantial hydrogen bonding on one of the phosphate groups. The half with the other phosphate is exposed freely to the exterior (Figure 5a,b). On the knowledge-based protein-ligand interaction analysis, 6 hydrogen bonds were observed on the phosphate group from Ser216, Ile219, Gly220, and Arg221, and 5 hydrophobic contacts were expected on Tyr46, Phe182, and Ala217 with the aliphatic chain in the middle part of the ligand structure. ChEMBL1161222 is structurally symmetric, and it is highly important to examine whether InteractionNet properly distinguishes the two phosphate groups in different surroundings. The heat map for the obtained atomic contributions of ChEMBL1161222 from the trained InteractionNet, arguably, showed a high correlation between human understanding and the machine-provided explanation (Figure 5c). InteractionNet focused on only one phosphate group, which resided inside the protein pocket, and predicted its high contribution to the increase in binding affinity. The influence of hydrophobic contacts was not observed in the heat map of 1KAV.

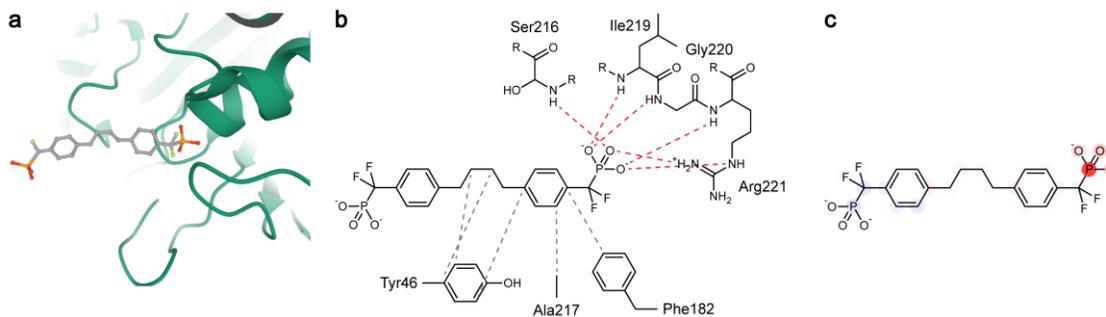

**Figure 5.** (a) Three-dimensional structure of the protein-ligand complex, 1KAV. The protein is depicted in a cartoon (green), and the ligand is depicted in color-coded ball-and-stick. Atom colors: gray (carbon), red (oxygen), orange (phosphorus), and light green (fluorine). (b) Knowledge-based estimation of protein-ligand interactions. Hydrogen bonds are depicted in red dashed lines, and hydrophobic contacts are depicted in gray dashed lines. (c) Heat map for the atomic contributions on the $K_d$ prediction, obtained from the LRP. The contributions are illustrated with color intensity of red (positive influence), white (zero influence), and blue (negative influence) colors.

*PDB 3F7H: Baculoviral IAP repeat-containing protein 7 with an azabicyclooctane-based antagonist (ChEMBL479725).* ChEMBL479725 can be divided into two parts by azabicyclooctane, the amide chain with one secondary amine, and the diphenylacetamide group. On the knowledge-based analysis, the amine and amide parts bound to 3F7H by four hydrogen bonds with their carboxyl and amide groups, and the diphenylacetamide group did not have interactions, except for one hydrophobic contact (Figure 6a,b). When compared to the machine-provided heat map, InteractionNet showed a highly positive focus on the terminal amine that participated in two hydrogen bonds (Asp138 and Glu143) and a little positive focus on the azabicyclooctane ring that participated in two hydrophobic contacts with the indole (Try147) and isobutyl (Leu131) groups. However, the diphenylacetamide group was predicted to slightly decrease the $K_d$ value (Figure 6c). The amide groups in the azabicyclooctane ring and the diphenylacetamide group had a negligible contribution to $K_D$, which concurred with the knowledge-based observation.

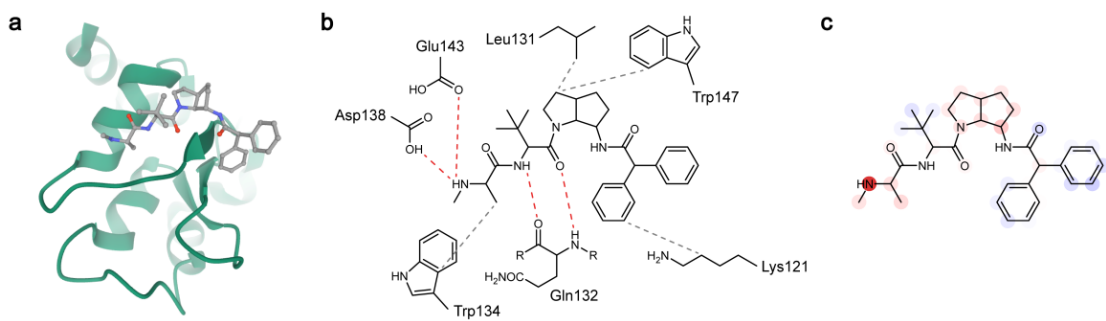

**Figure 6.** (a) Three-dimensional structure of the protein-ligand complex, 3F7H. The protein is depicted in a cartoon (green), and the ligand is depicted in color-coded ball-and-stick. Atom colors: gray (carbon), red (oxygen), and blue (nitrogen). (b) Knowledge-based estimation of protein-ligand interactions. Hydrogen bonds are depicted in red dashed lines, and hydrophobic contacts are depicted in gray dashed lines. (c) Heat map for the atomic contributions on the prediction of $K_d$, obtained from the LRP. The contributions are illustrated with color intensity of red (positive influence), white (zero influence), and blue (negative influence) colors.

*PDB 4IVB: Tyrosine-protein kinase JAK1 with an imidazopyrrolopyridine-based inhibitor (ChEMBL2386633).* In the 4IVB complex, ChEMBL2386633 resided in between the two lobes of JAK1 and was expected to have four hydrogen bonds, i.e., two in the imidazopyrrolopyridine group and two in the hydroxyl group and three hydrophobic contacts with JAK1 (Figure 7a,b). The ChEMBL2386633 heat map showed a similar contribution pattern, predicting a highly positive contribution from the nitrogen atoms of imidazopyrrolopyridine and one oxygen atom. The most-focused atom was the oxygen of the hydroxyl group, which participated in two hydrogen bonds with Ser963 and Glu966 of JAK1. The four nitrogen atoms in imidazopyrrolopyridine were given a high contribution, but only two nitrogen atoms participated in the hydrogen bond. The prediction on the cyanocyclohexyl group was not influential to the $K_d$, which corresponded with the 3D structure showing the exposure of the group to the exterior.

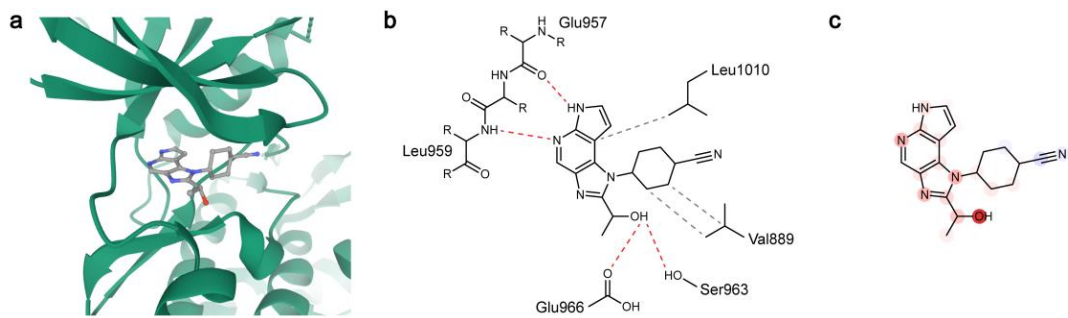

**Figure 7.** (a) Three-dimensional structure of the protein-ligand complex, 4IVB. The protein is depicted in a cartoon (green), and the ligand is depicted in color-coded ball-and-stick. Atom colors: gray (carbon), red (oxygen), and blue (nitrogen). (b) Knowledge-based estimation of protein-ligand interactions. Hydrogen bonds are depicted in red dashed lines, and hydrophobic contacts are depicted in gray dashed lines. (c) Heat map for the atomic contributions on the prediction of $K_d$, obtained from the LRP. The contributions are illustrated with color intensity of red (positive influence), white (zero influence), and blue (negative influence) colors.

# Conclusions

In conclusion, we presented a graph neural network (GNN) that modeled the noncovalent (NC) interactions and discussed the in-depth analysis of the model combined with the explainability technique for understanding deep-learning prediction. In the graph-based deep-learning models, there has been less attention to the NC interactions compared with the bonded interactions because of the ambiguity of NC connectivity and its effectiveness over the traditional covalent-bond-based strategies. InteractionNet, presented herein, showed satisfactory predictive-ability for predicting the dissociation constant with RMSE of 1.321 on the PDBbind v2018 dataset. The NC convolution layers enhanced InteractionNet's prediction accuracy, even without the utilization of the traditional bonded connectivity. We further demonstrated that InteractionNet successfully captured the important NC interactions between a protein and a ligand from a given complex through posthoc LRP analysis. The visualization of the atomic contributions showed a strong correlation with the actual hydrogen bonds in the complex. In the case of the ligand that had multiple hydrogen-bond donors and acceptors, the positive atomic contributions were observed only on the atoms participating in the actual hydrogen bonds. We believe that our model would widen the applicable tasks of the chemical, deep-learning models to the problems beyond the bonded interactions within a single molecule and also provide a meaningful explanation for the prediction, enabling the real-world applications that require prediction evidence and reliability.

# Methods

**Dataset.** We employed the PDBbind v2018 dataset for the evaluation target of our InteractionNet models.[23,24] We used the refined set from the provided dataset, consisting of 4,462 protein-ligand complexes with their experimentally measured $K_d$ values. Initially, all protein-ligand data were loaded by RDKit 2019.09.2[35] and Openbabel 3.0.0[36], and inspected for improper conformation. During the inspection process, the molecules that failed for loading or contained atomic collisions (interatomic distance below 1 Å) were excluded from the dataset. After inspection, 4,186 protein-ligand complexes were obtained. The protein structure was cropped by retrieving the atoms of a protein within the range cutoff (3, 4, 5, or 6 Å), and the size of the protein-ligand complex structure was reduced for faster training. Only heavy atoms were considered in the entire preparation. Atomic features for building the feature matrix are listed in Table S1. In each cross-validation experiment, the refined set was randomly split into a training set, a validation set, and a test set, on an 8:1:1 ratio. Twenty results were obtained through 20-fold cross-validation, and the averaged results were reported.

**Network Training and Evaluation.** All models were implemented by using TensorFlow 2.0.0[37] on Python 3.6.9. The training was controlled by learning-rate scheduling, early-stopping techniques, and gradient norm scaling. The learning rate was initially set to 0.00015 and lessened by a factor of 0.75 when the validation loss did not decrease within the previous 200 epochs, and the termination proceeded when the loss stopped decreasing for the previous 400 epochs. To avoid gradient exploding, a clipping parameter of 0.5 was used for gradient norm scaling. For the loss function, mean-squared-error (MSE) was used and optimized by the Adam optimizer.[38] The list of hyperparameters explored is described in Table S2. All experiments were conducted on an NVIDIA GTX 1080Ti GPU, an NVIDIA RTX 2080Ti GPU, or an NVIDIA RTX Titan GPU.

**Layer-Wise Relevance Propagation (LRP).** We performed the LRP as a post-modeling explainability method. Three LRP rules, LRP-0, LRP-ε, and LRP-γ, were used for the calculation of relevance on each layer from the trained model. We adopted the LRP-0 for the output layer, LRP-ε for the FC layers, and the LRP-γ for the $CV_{[C]}$ and $CV_{[NC]}$ layers, based on the guideline described elsewhere.[31,32] Obtained relevance for the atomic feature was reduced to the atomic contribution by summation across features. The graph-embedding layers were omitted for the relevance calculation, because the graph-embedding layers only redistributed the relevance between features, not between atoms, resulting in the same atomic contribution before and after redistribution. For the parameters ε and γ, 0.25 was used for all LRP-ε, and 100 was used for all LRP-γ layers. The cross-validation trial that was most similar to the average in root-mean-squared-error (RMSE) was used for LRP analysis, and the LRP examples were chosen from the test set of the trial, which were predicted accurately, for comparison with knowledge-based analysis. Three-dimensional visualization of the molecular structure was obtained by Mol*,[40] and the expected hydrogen bonds and hydrophobic contacts were determined by the rules RCSB PDB use.[41-44]

# Acknowledgments

This work was supported by the KAIST-funded AI Research Program for 2019.

# Author contributions

H.C., E.K.L., and I.S.C. developed the concept, and H.C. constructed the deep-learning architectures and performed the experiments. H.C. wrote the manuscript, and E.K.L. and I.S.C. supervised the work and reviewed the manuscript.

## Competing interests

The authors declare no competing interests.

## Additional information

**Correspondence** and requests for materials should be addressed to E.K.L. or I.S.C.

SUPPLEMENTARY INFORMATION

# InteractionNet: Modeling and Explaining of Noncovalent Protein-Ligand Interactions with Noncovalent Graph Neural Network and Layer-Wise Relevance Propagation

Hyeoncheol Cho[1], Eok Kyun Lee[1*], and Insung S. Choi[1*]

[1]Department of Chemistry, KAIST, Daejeon 34141, Korea.
Email: eklee@kaist.ac.kr, ischoi@kaist.ac.kr.

## CONTENTS

- **Experimental Section**.
- **Table S1.** Atom features used in the graph representation of molecules.
- **Table S2.** Hyperparameters explored for InteractionNet.
- **Figure S1.** (Left) Scatterplots and (right) error distributions of predicted dissociation constants and experimental values included in the test set across the best, averaged, and worst cross-validation trials.

**Table S1. Atom features used in the graph representation of molecules.**

| Feature | Description | Type | Size |
| --- | --- | --- | --- |
| Atom type | atom type | one-hot | 24 |
| Atomic number | atomic number | integer | 1 |
| Degree | the number of heavy atom neighbors (0 to 6) | one-hot | 7 |
| Number of hydrogens | the number of neighboring hydrogens (0 to 4) | one-hot | 5 |
| Implicit valence | the number of implicit hydrogens (0 to 6) | one-hot | 7 |
| Hybridization | $sp$, $sp^2$, $sp^3$, $sp^3d$, or $sp^3d^2$. | one-hot | 5 |
| Formal charge | atomic formal charge (-3 to +3) | one-hot | 7 |
| Ring size | whether this atom belongs to a ring (ring size: 3 to 8) | binary | 6 |
| Aromaticity | whether this atom is part of an aromatic system. | binary | 1 |
| Acid/base | whether this atom is acidic or basic | binary | 2 |
| Hydrogen bonding | whether this atom is a hydrogen bond donor or acceptor | binary | 2 |
| Total | | | 67 |

**Table S2. Hyperparameters explored for InteractionNet.**

| Group | Hyperparameter | Size |
|---|---|---|
| Graph Embedding | number of output units | 128, 256, 512 |
| | number of embedding layers | 1, 2 |
| Graph Convolution | number of output units | 128, 256, 512 |
| | number of convolution layers each | 0, 1, 2, 3 |
| Fully connected | number of output units | 128, 256, 512 |
| | number of fully-connected layers | 2, 3 |
| | l2 regularization | 0.0025, 0.005, 0.01, 0.02, 0.04 |
| Training | batch size | 32 |
| | initial learning rate | 0.00015 |
| | patience | 100, 200, 400 |
| | loss | MSE |
| | gradient descent method | Adam |

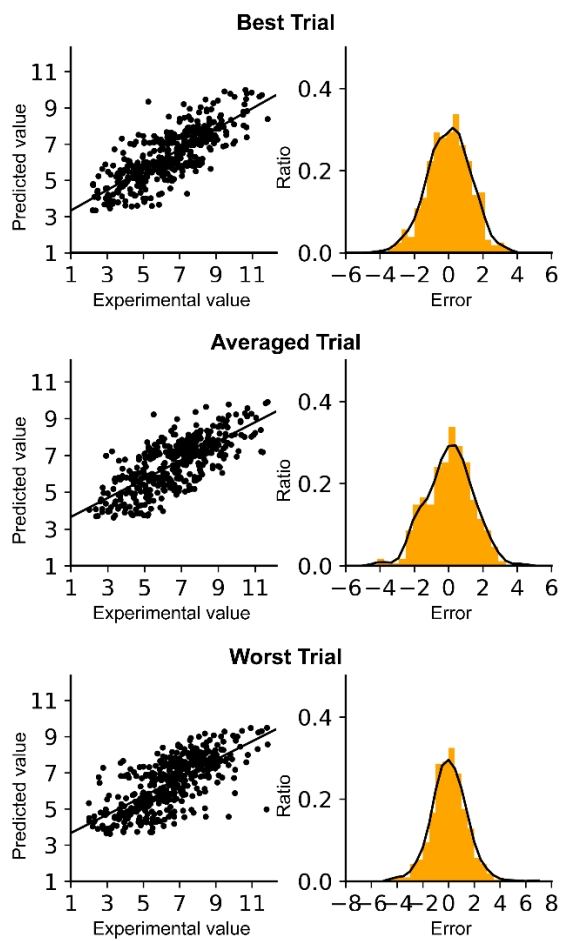

**Figure S1.** (Left) Scatterplots and (right) error distributions of predicted dissociation constants and experimental values, included in the test set across the best, averaged, and worst cross-validation trials. (Left) The scatterplot for predicted versus experimental constants is depicted with the solid trend line. (Right) The histogram (orange) and distribution (black) for predictions from the test set.